\documentclass[twocolumn,prd,nofootinbib,aps,prl,floats,floatfix,amsmath,amssymb,longbibliography,secnumarab
ic]{revtex4-1} %
\usepackage[final]{graphicx}
\usepackage{hyperref}
\usepackage{amsmath}
\usepackage{bbm}
\usepackage{amsfonts}
\usepackage{amssymb}
\usepackage{latexsym}
\usepackage{graphicx}
\usepackage[english]{babel}
\usepackage{multirow}
\usepackage{float}
\usepackage{url}
\usepackage{slashed}
\usepackage{xcolor} 
\usepackage[utf8]{inputenc}
\usepackage{bbding}
\usepackage[nice]{nicefrac}
\usepackage{verbatim}
\usepackage{selinput}
\SelectInputMappings{adieresis={ä},germandbls={ß},Euro={€},} 

\def\sfrac#1#2{{\textstyle{#1\over #2}}}
\newcommand{\be}{\begin{equation}}
\newcommand{\ee}{\end{equation}}
\newcommand{\ba}{\begin{array}}
\newcommand{\ea}{\end{array}}
\newcommand{\bea}{\begin{eqnarray}}
\newcommand{\eea}{\end{eqnarray}}
\newcommand{\sss}{\scriptscriptstyle}

\newcommand{\nn}{\nonumber}

\newcommand{\R}{{\sss R}}
\newcommand{\BR}{{\rm BR}}

\newcommand{\ct}{c_\theta}
\newcommand{\st}{s_\theta}
\newcommand{\napprox}{\not\approx}

\begin{document}

\title{Pseudo-Goldstone dark matter confronts cosmic ray
and collider anomalies}

\author{James M.\ Cline}
\author{Takashi Toma}
\affiliation{McGill University, Department of Physics, 3600 University St.,
Montr\'eal, QC H3A2T8 Canada}
\begin{abstract}

Persistent excesses in the spectra of gamma rays from the galactic
center and cosmic ray antiprotons  can be explained by dark matter of
mass  $50\sim 65$\,GeV annihilating into $b$ quarks, but this is
typically hard to reconcile with direct detection constraints.  We
resolve this tension using a simple class of models, where dark matter is
a  pseudo-Nambu-Goldstone boson, having naturally momentum-suppressed
couplings to nuclei.  Exploring the parameter space of the model, we
find that it can explain the cosmic ray anomalies while remaining
compatible with constraints from the relic abundance and annihilation
in dwarf spheroidal galaxies. In certain regions of parameter space,
the Higgs-dark matter coupling can help stabilize the Higgs potential
up to the Planck scale.  The scalar partner of the dark matter is
an extra Higgs boson, that 
can explain a tentative diphoton excess observed by CMS, and 
an excess of $b\bar b$ signal
from LEP, if its mass is $\sim 96$\,GeV and the 
model is extended to include a heavy scalar quark.  This extended
model predicts a monochromatic gamma-ray line near 64\,GeV,
at a level close to current experimental sensitivity, from 
dark matter annihilations in the galaxy.

\end{abstract}
\maketitle

\section{Introduction}

Although there is so far no clear evidence for nongravitational
interactions of dark matter (DM), it has been suggested that excess of
GeV-scale gamma rays from the galactic center (GC), observed by the
Fermi-LAT~\cite{TheFermi-LAT:2015kwa, TheFermi-LAT:2017vmf}, could be
due to DM annihilations~\cite{Goodenough:2009gk, Hooper:2010mq,
Hooper:2011ti,Hooper:2013rwa,Daylan:2014rsa,
Calore:2014xka,Abazajian:2014fta}.  A leading alternative astrophysical
explanation is that the excess
comes from an unresolved population of millisecond pulsars or other
astrophysical
sources~\cite{Mirabal:2013rba,Calore:2014oga,OLeary:2015qpx,
Bartels:2018xom,
Brandt:2015ula}, although arguments against the pulsar hypothesis
have been presented~\cite{Hooper:2013nhl,Cholis:2014lta,Haggard:2017lyq}.  Until such possible point sources
are resolved, the DM hypothesis
remains viable.  In particular, ref.~\cite{Leane:2019xiy} has
recently called into question  claims that the statistical properties
of the GeV excess are more consistent with unresolved point sources
than with DM annihilations~\cite{Bartels:2015aea,Lee:2015fea}, rekindling the motivation to consider
particle physics models for the GC excess.

At the same time, the AMS collaboration obtained the first 
precise measurement of the spectrum of 
cosmic ray antiprotons~\cite{Aguilar:2016kjl}.  It has been
analyzed by numerous groups who find an excess over the expected
flux at energies of 10-20\,GeV~\cite{Fujita:2009wk, Bringmann:2014lpa, Cirelli:2014lwa,
Hooper:2014ysa, Kohri:2015mga, Cuoco:2016eej, Cui:2016ppb}.  Intriguingly, this signal can 
be simultaneously explained with the GC gamma-ray excess by 
DM annihilating into $b\bar b$~\cite{Cholis:2019ejx, Lin:2019ljc, Carena:2019pwq}.  
In this paper we propose that both the GC and the $\bar p$ excesses
can be explained by the annihilation of pseudo-Nambu Goldstone dark
matter (pNGB DM).

A natural explanation for why DM ($\chi$) should annihilate mainly
into $b\bar b$ is the Higgs portal, since $h\to b\bar b$ is the 
dominant decay channel of the Higgs boson.  However this tends to be
strongly ruled out by direct detection constraints~\cite{Aprile:2018dbl}, because $h$
couples relatively strongly to nucleons~\cite{Casas:2017jjg}.  For
this reason, singlet scalar DM in the mass range of interest is
excluded except for  $m_\chi \in (55-62.5)$\,GeV, just below
the Higgs resonance \cite{Cline:2013gha,Athron:2017kgt}.  But for
$m_\chi < m_h/2$, the cross section for $\chi\chi\to b\bar b$ at 
present times is highly suppressed compared to that during the
time of freeze-out in the early universe (as we will discuss below),
making singlet scalar DM unsuitable for explaining the cosmic ray
excesses.  

One way of circumventing the direct detection constraints for
Higgs portal models is by taking the DM to be
fermionic, coupling to $b$ quarks through a pseudoscalar such that
DM-nucleon interactions are spin-dependent and
velocity-suppressed~\cite{Ipek:2014gua,Escudero:2016kpw,
Abe:2018emu,Boehm:2014hva, Arina:2014yna}.  Here we will use a different
strategy, taking advantage of an approximate symmetry, following 
refs.~\cite{Barger:2008jx,Barger:2010yn,Gross:2017dan}.  
There a simple class of models was considered
where the required cancellation for direct detection occurs as
consequence of a softly broken global U(1) symmetry, 
rather than through fine tuning.
The DM candidate is the pseudo-Nambu
Goldstone boson arising from spontaneous breaking of
the global U(1) symmetry, with Higgs portal couplings.

In this case, the pNGB retains the feature of being derivatively
coupled with other particles. Its interaction with matter is therefore
highly suppressed for nonrelativistic DM, as in direct detection
experiments. Loop corrections do not have this feature because of the
explicit breaking of the  global U(1) symmetry.  This has been studied
in ref.~\cite{Azevedo:2018exj, Ishiwata:2018sdi}, where it was
found that loop corrections are nevertheless very small and well within
current direct detection bounds.

The pNGB dark matter model is quite economical, with only four free parameters:
the DM mass $m_\chi$, the mass of its scalar partner (a second Higgs
boson $h_2$), the mixing angle $\theta$ between $h_2$ and the observed
Higgs $h_1$, and the vacuum expectation value (VEV) $v_s$ that
spontaneously breaks the U(1) symmetry (or alternatively the quartic coupling
$\lambda_S$ of the new complex scalar field $S$).
In ref.~\cite{Gross:2017dan}, it was shown that
one can easily satisfy the requirements of getting the correct relic
density, while respecting other constraints like the Higgs invisible
width and perturbative unitarity.
Moreover, probing the pNGB DM at the LHC has been discussed in
ref.~\cite{Huitu:2018gbc}, and further generalizations of the model 
have been considered~\cite{Alanne:2018zjm, Karamitros:2019ewv}.
However its implications for indirect detection have not been
considered prior to this work.

A possible bonus of the pNGB DM model 
is that it 
can enhance the stability of the Higgs quartic coupling to higher
renormalization scales, in contrast to the SM prediction that
the Higgs quartic coupling should become negative around
$10^{11}\,$GeV~\cite{Degrassi:2012ry}, triggering an instability. 
This is a consequence of the Higgs portal coupling, that affects the
renormalization group evolution of the Higgs quartic
coupling~\cite{Lebedev:2012zw, EliasMiro:2012ay,Khoze:2014xha,
Falkowski:2015iwa,Athron:2018ipf}.  
We identify regions of parameter space where the instability can be
avoided and the quartic couplings are perturbative up to the Planck
scale, while satisfying the relic abundance constraints and accounting for the
cosmic ray excesses. 

A further set of tentative anomalies has been identified in  results
coming from the LEP and CMS experiments, that both point to a new
Higgs-like particle $h_2$ at a mass close to $96\,$GeV. The LEP anomaly is
a mild excess  $b\bar b$ pairs that would arise from decays of
$h_2$ following $e^+e^- \to Z h_2$ production via
Higgs-strahlung~\cite{Barate:2003sz}. At a similar mass, CMS observes a
2.9\,$\sigma$ local excess in the diphoton channel, that could come from $gg\to
h_2\to\gamma\gamma$~\cite{CMS:2015ocq, CMS:2017yta,Sirunyan:2018aui}.
These observations have motivated
theorists to propose a number of
models~\cite{Cao:2016uwt,Fox:2017uwr,Haisch:2017gql,Liu:2018xsw,Vega:2018ddp,LiuLiJia:2019kye,Domingo:2018uim,Biekotter:2019kde,Biekotter:2019mib,Biekotter:2017xmf} 
A necessary
ingredient of the pNGB DM model is an additional scalar $h_2$ that
mixes with the SM Higgs, denoted as $h_1$.   It is possible to
take $m_{h_2} = 96\,$GeV while still being consistent with the cosmic
ray excesses.  Although the LEP
observation could come directly from the pNGB DM  setup, given large
enough mixing between the two scalars, the CMS excess requires an
additional charged and colored scalar $\Phi$ coupling to the singlet.  We
make a systematic study of the potential for such extensions of the
model to simultaneously explain both tentative signals. 
 as well as
the collider anomalies.  
We show that the extended model including the colored
scalar provides an additional means of discovering pNGB DM, by its
annihilation to monochromatic gamma rays in the galaxy.

The paper is organized as follows.  
We review the framework of pNGB DM in section~\ref{review} and then show  in
section~\ref{numerical} that it can account for the anomalous cosmic
ray signals in gamma rays and antiprotons, for DM mass near $m_\chi
\sim 65\,$GeV, over wide ranges of the other free parameters.  
In section~\ref{sec:stability} we explore Higgs stability and perturbativity,
showing that it can be achieved if $m_{h_2}\gtrsim 140\,$GeV.
Section~\ref{sec:lep-cms} analyzes a broad class of models that can
potentially explain the LEP and CMS anomalies, singling out one that is most 
promising. We summarize and give conclusions in section~\ref{conclusion}.

\section{pNGB dark matter}
\label{review}

We begin by reviewing the particle physics model, which has just one
additional complex scalar field $S = (v_s + s + i\chi)/\sqrt{2}$ relative
to the standard model (SM). The scalar potential is given by
\begin{align}
\mathcal{V} &= -\frac{\mu_H^2}{2}|H|^2 + \frac{\lambda_H}{2}|H|^4-\frac{\mu_S^2}{2}|S|^2 + \frac{\lambda_S}{2}|S|^4
\nonumber\\
 &~~~+\lambda_{HS}|H|^2|S|^2 - \frac{m_\chi^2}{4}\left(S^2 + {S^*}^2\right),
\end{align}
where the last term explicitly breaks the global U(1) symmetry $S\to
e^{i\alpha}S$, giving mass $m_\chi$ to the NGB DM candidate
 $\chi$.
This explicit breaking term can be derived from an extended model with
gauged $U(1)$ symmetry at high energy scale~\cite{Gross:2017dan}.
From the viewpoint of technical naturalness, there is no need for
the soft-breaking mass to be small compared to $\mu_S^2$; moreover
the model's mechanism for suppressing direct detection signals does 
not require any hierarchy between the two mass scales.

 With no loss of generality, $m_\chi^2$ can be made real-valued
by absorbing its phase with a field redefinition of
 $S$. Then the DM $\chi$ remains stable
 due to a $\mathbb{Z}_2$ symmetry even after  $S$ gets its VEV. 
 The field $H$ is the SM Higgs doublet, $H=(0,(v+h)/\sqrt{2})^T$. 
The VEVs $v_s$ and $v$ are
\begin{equation}
 \left(
  \begin{array}{c}
  v^2\\ v_s^2\end{array}
 \right) = \frac{1}{\lambda_H\lambda_S-\lambda_{HS}^2}
 \left(
 \begin{array}{cc}
  \lambda_S & -\lambda_{HS}\\
  -\lambda_{HS} & \lambda_H
 \end{array}\right)
 \left(
  \begin{array}{c}
   \mu_H^2 \\
   \mu_S^2+m_\chi^2
  \end{array}
	\right).
\end{equation}
The mass matrix for $h$ and $s$ can be written as 
\begin{equation}
 M^2 = \left(
	 \begin{array}{cc}
	  \lambda_H v^2 & \lambda_{HS}v v_s\\
	  \lambda_{HS} v v_s & \lambda_S v_s^2
	 \end{array}
	\right)\,,
\end{equation}
which is diagonalized by the rotation 
\begin{equation}
 \left(
  \begin{array}{c}
   h \\
   s
  \end{array}
 \right)= \left(
 \begin{array}{cc}
  \ct & \st \\
  -\st & \ct
 \end{array}
	  \right)
 \left(
  \begin{array}{c}
   h_1\\
  h_2\end{array}
 \right),
\end{equation}
with $\ct = \cos\theta$, $\st=\sin\theta$, and the mixing angle given by
\begin{equation}
 \tan 2\theta = \frac{2\,\lambda_{HS} v v_s}{\lambda_S v_s^2 - 
  \lambda_H v^2}\,.
\end{equation}
The mass eigenstate $h_1$ is identified as the SM-like Higgs boson with
mass $m_{h_1}=125~\mathrm{GeV}$, and $h_2$ is a new $CP$-even
state. 
The trilinear couplings of the DM to the $h_i$ bosons
are expressed in terms of $\theta$ and the masses as
\begin{equation}
 \mathcal{V}\supset
  -\frac{\st\, m_{h_1}^2}{2v_s}h_1\chi^2
  +\frac{\ct\, m_{h_2}^2}{2v_s}h_2\chi^2,
 \label{dmhc}
\end{equation}
while the $h_i$ couplings to SM fermions $f$ have the form
\begin{equation}
 \mathcal{L}\supset
  -\frac{m_f}{v}\left(\ct\, h_1+\st\, h_2\right)
\overline{f}f\ .
\label{fhc}
\end{equation}

From the structure of Eqs.~(\ref{dmhc}) and (\ref{fhc}), one can see that the 
matrix element for $\chi f\to\chi f$
scattering mediated by $h_{1,2}$ exchange exactly vanishes at zero 
momentum transfer, which is the
key feature for making the model safe from direct detection
constraints.  On the other hand for the annihilation $\chi\chi\to
f\overline{f}$, the cancellation is ineffective as long as
$m_{h_1}\napprox m_{h_2}$, since the $s$-channel propagator 
carries significant momentum.
The main annihilation channel determining the DM relic abundance is
$\chi\chi\to b\overline{b}$ for $5~\mathrm{GeV}\lesssim m_\chi\lesssim
70~\mathrm{GeV}$, covering the mass range that is relevant 
for the present
work.\footnote{See the next section for discussion of the
subdominant $\chi\chi \to WW$ annihilation channel.}
The DM annihilation cross section for $\chi\chi\to b\bar
b$ is given by
\begin{align}
 \sigma_{b\overline{b}} v_{\rm rel} &\approx \frac{\ct^2\,\st^2\,
 m_b^2}{4\pi v^2 v_s^2}\left(1-4\frac{m_b^2}{m_\chi^2}\right)^{3/2}\label{sigv}\\
 &~~~\times\left|\frac{m_{h_1}^2}{s-m_{h_1}^2+im_{h_1}\Gamma_{h_1}}
 -\frac{m_{h_2}^2}{s-m_{h_2}^2+im_{h_2}\Gamma_{h_2}}\right|^2,
 \nonumber
\label{sigv}
\end{align}
where $\sqrt{s}$ is the total DM energy in the center of mass frame, and
$\Gamma_{h_i}$ is the total decay width of $h_i$.
For quantitative determination of the DM relic abundance, we use the full
thermally averaged annihilation cross section 
$\langle\sigma v_{\rm rel}\rangle$ computed using MicrOmegas~\cite{Belanger:2018mqt}.

\begin{figure}[t]
\begin{center}
 \includegraphics[scale=0.6]{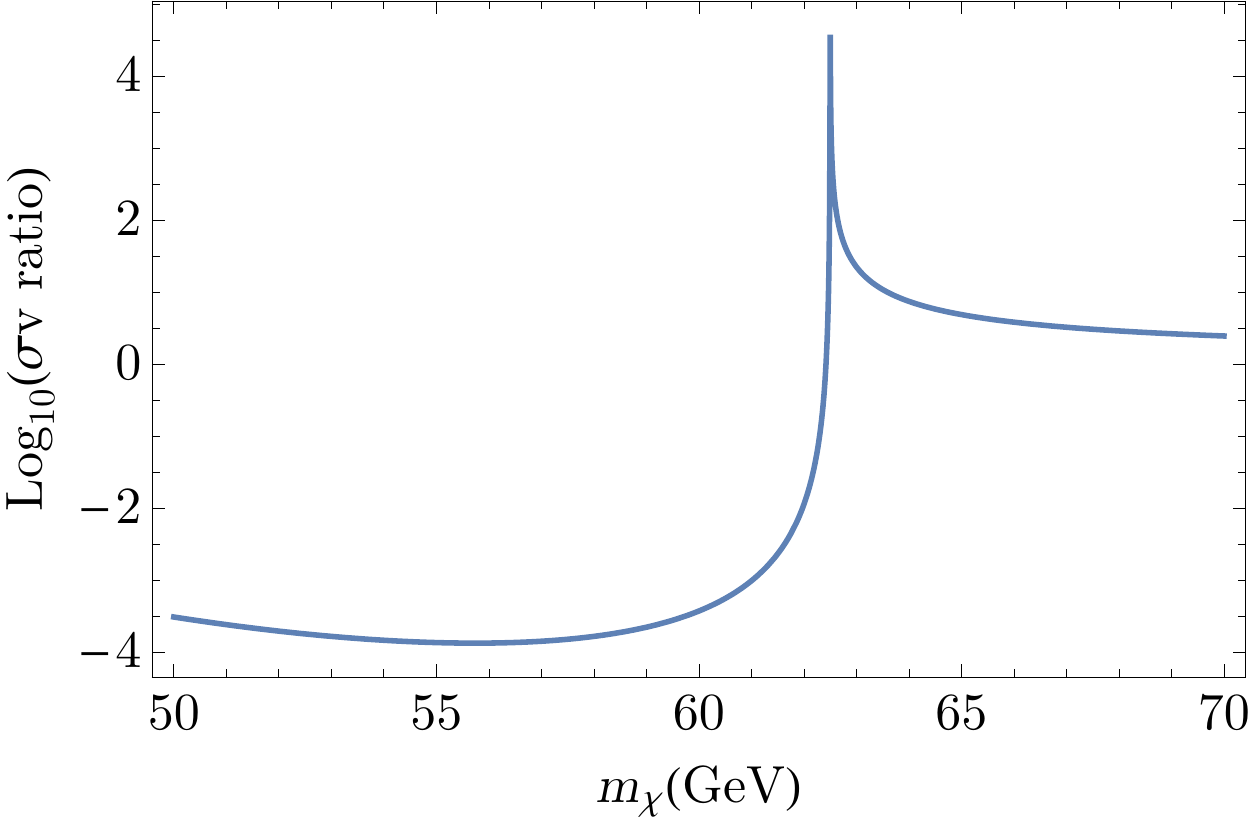}
 \caption{Ratio of $\langle\sigma_{b\bar b} v_{\rm rel}\rangle$ at present times to that in the early universe, taking the
freezeout temperature to be $T=m_\chi/20$.}
 \label{fig:ratio}
\end{center} 
\end{figure}

On the other hand for annihilation within our galaxy, Eq.~(\ref{sigv})
at $s=4m_\chi^2$ provides a very good approximation in the DM mass
region we consider because the DM is very slowly moving, $v\sim
10^{-3}\,c$, and its kinetic energy is much smaller than the Higgs
decay width,  $(s-4m_\chi^2)/m_\chi \sim m_\chi v^2\ll \Gamma_{h_i}$.
Therefore the
phase space averaging has a negligible effect on Eq.~(\ref{sigv})
even exactly on the resonance $4 m_\chi^2 = m_{h_i}^2$,   and
we can use this expression for predicting indirect detection signals.

It is interesting that Eq.~(\ref{sigv}) can differ significantly from 
$\langle\sigma_{b\bar{b}} v_{\rm rel}\rangle$ in the early universe near the
resonance at $m_\chi = m_{h_i}/2$~\cite{Ibe:2008ye}.  We illustrate
this in figure \ref{fig:ratio} by plotting the ratio of 
$\langle\sigma_{b\bar{b}} v_{\rm rel}\rangle$ at late times versus at the time of
freezeout as a function of $m_\chi$.  This shows that the indirect
signal is suppressed by orders of magnitude for $m_\chi < m_h/2$ and
motivates our search for models with $m_\chi > m_h/2$, where the
indirect signal is moderately enhanced, but can nevertheless satisfy
Fermi dwarf spheroidal 
constraints~\cite{Ackermann:2015zua,Drlica-Wagner:2015xua,
Ahnen:2016qkx}.

If $m_\chi< m_{h_1}/2\approx 62.5~\mathrm{GeV}$, the trilinear coupling in
Eq.~(\ref{dmhc}) gives rise to an
invisible decay channel for the Higgs boson, $h_1\to\chi\chi$, whose
partial width is
\begin{equation}
 \Gamma_{\rm inv} = \frac{\st^2 \, m_{h_1}^3}{32\pi\,
  v_s^2}\sqrt{1 - 4\frac{m_\chi^2}{m_{h_1}^2}}\ .
\label{eq:inv}
\end{equation}
The branching ratio for invisible decays is
experimentally constrained at the level of $\mathrm{BR}_\mathrm{inv} <
0.19$ by the CMS Collaboration~\cite{Sirunyan:2018owy}, and $0.26$ by
the ATLAS Collaboration~\cite{Aaboud:2019rtt}.
Ignoring the phase space suppression (which is valid for $m_\chi
\lesssim 60\,$GeV), this gives the approximate constraint 
$\st/v_s \lesssim 0.2~{\rm\, TeV}^{-1}$.

\section{Fitting cosmic ray excesses}
\label{numerical}

\begin{figure*}[t]
\centerline{
 \includegraphics[width=0.45\hsize]{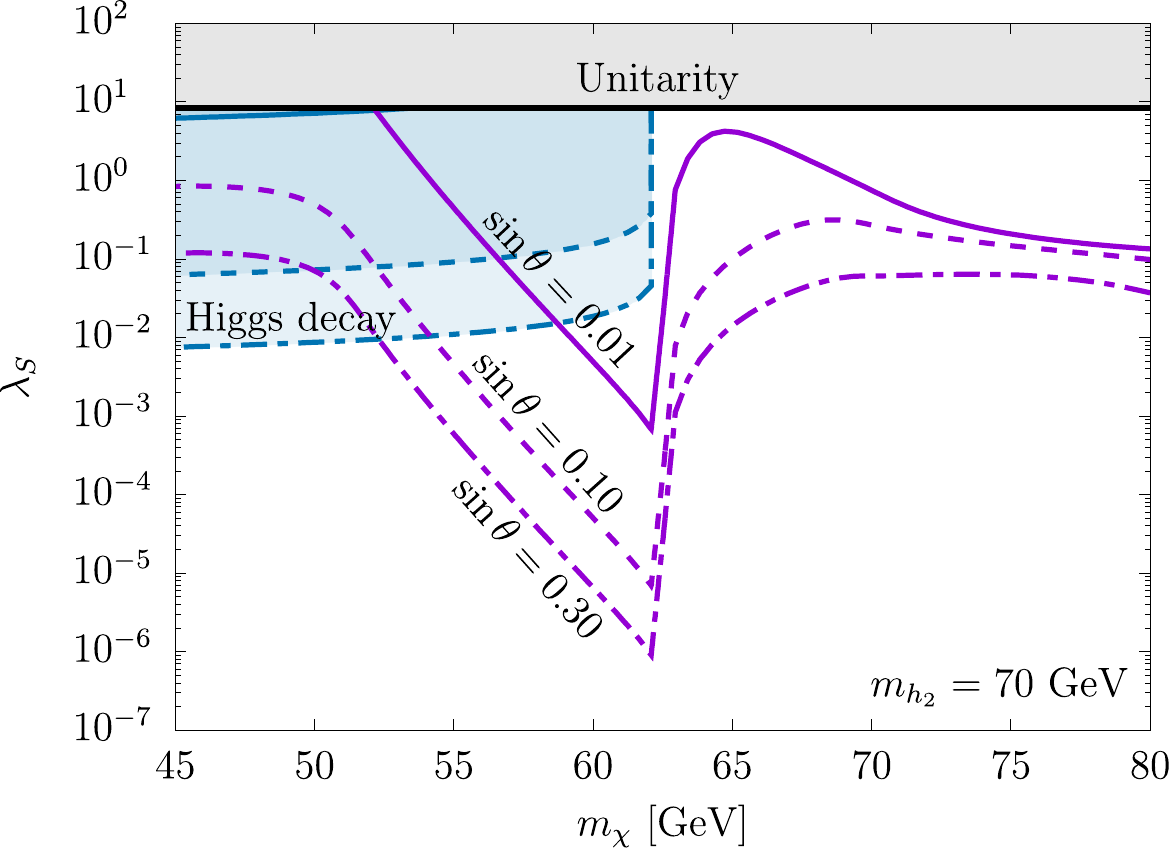}
 \hspace{0.5cm}
 \includegraphics[width=0.45\hsize]{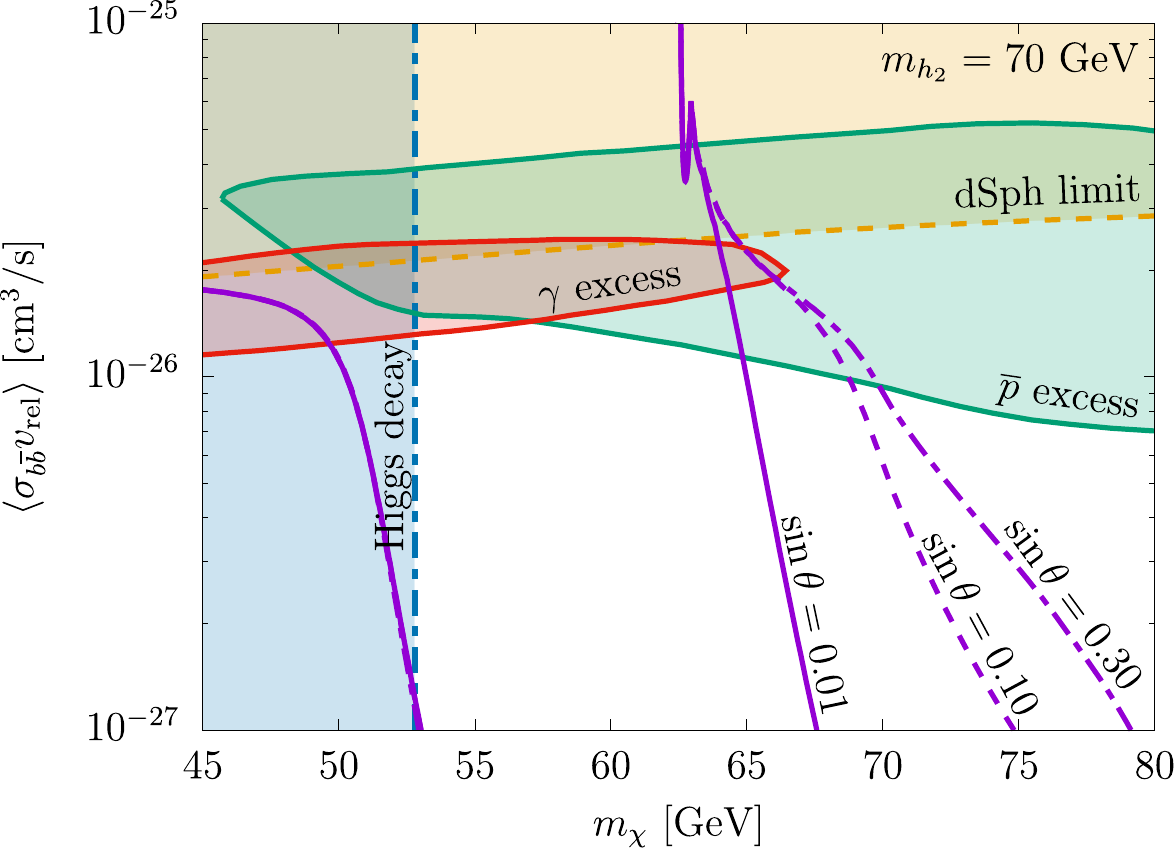}
 }
 \centerline{
 \includegraphics[width=0.45\hsize]{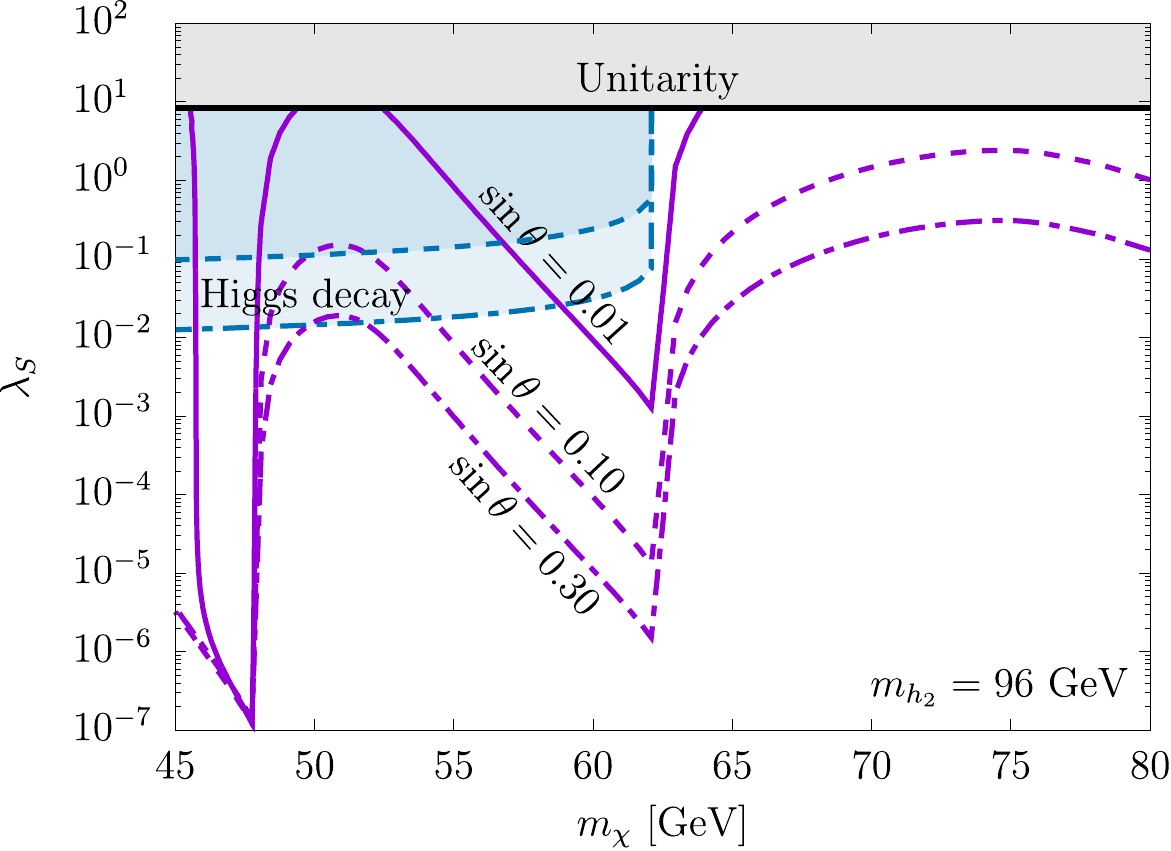}
  \hspace{0.5cm}
 \includegraphics[width=0.45\hsize]{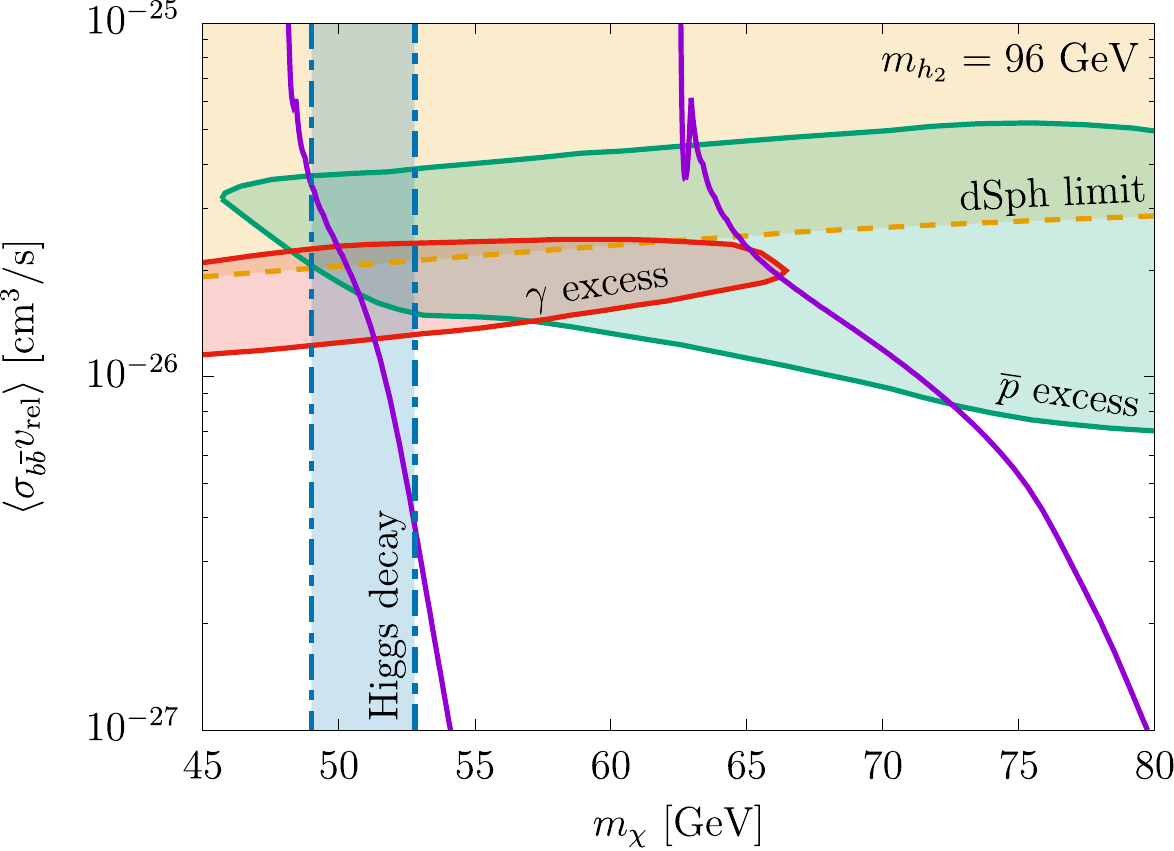}
 }
 \centerline{
 \includegraphics[width=0.45\hsize]{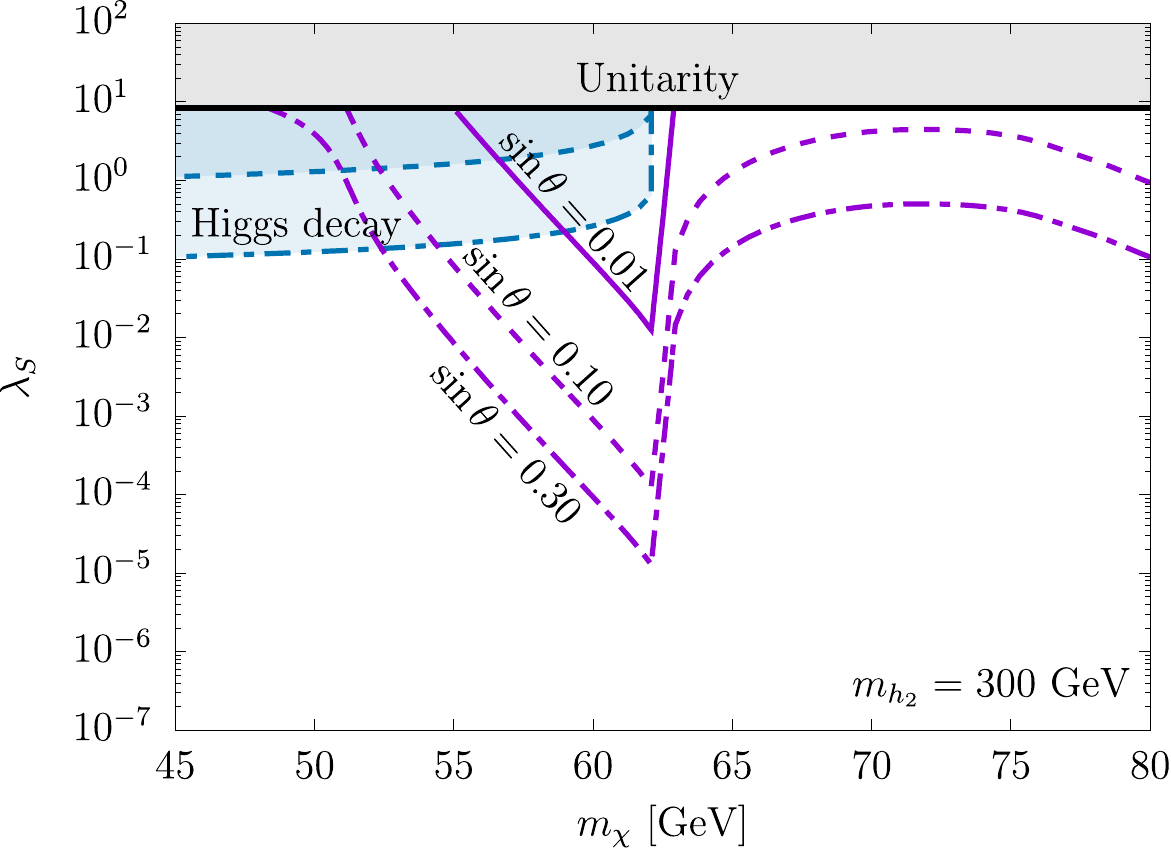}
  \hspace{0.5cm}
 \includegraphics[width=0.45\hsize]{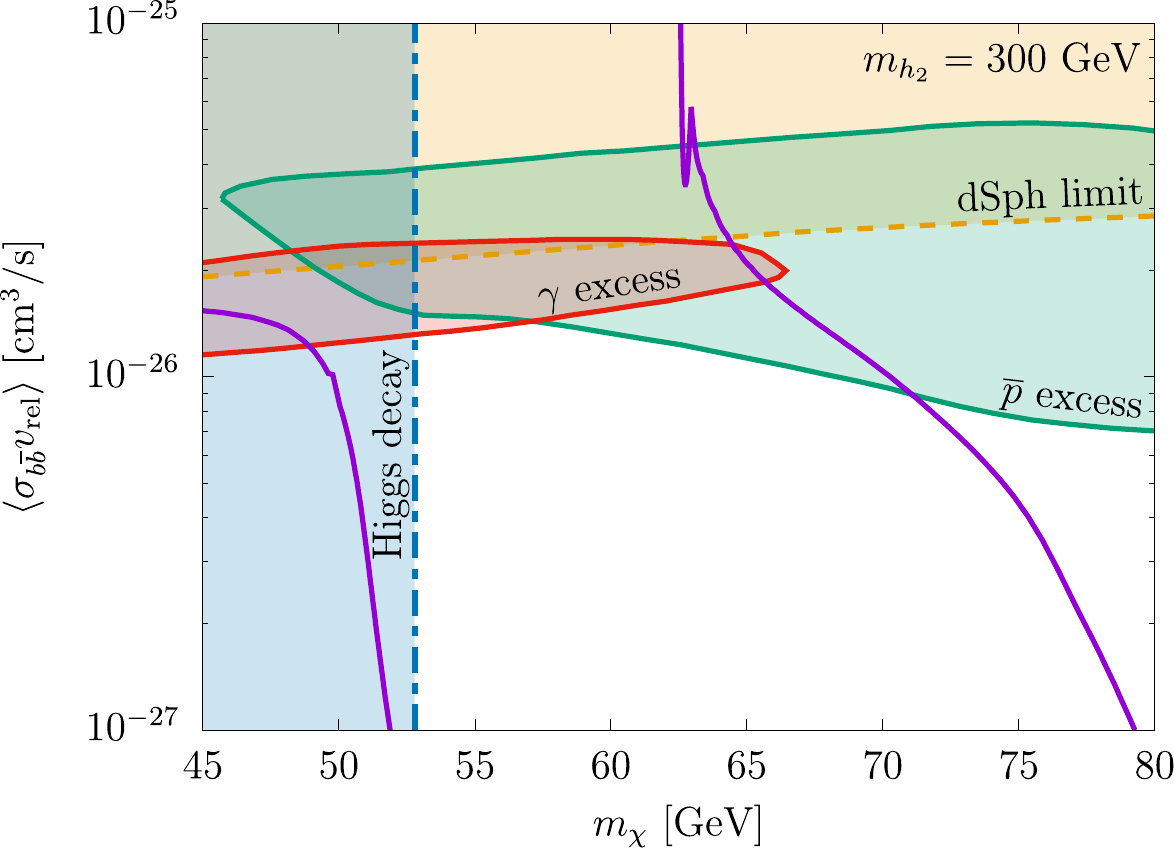}
}
\caption{Left: constraints from the thermal relic abundance (purple curves), 
Higgs invisible width
(blue region), and
perturbative unitarity (gray region, $\lambda_S>8\pi/3$).
The mixing angle $\sin\theta$ is varied between $0.01$ (solid), $0.1$
(dashed) and $0.3$ (dot-dashed).  The new $CP$-even Higgs mass is
$m_{h_2}=70$ GeV (top),
 $96$ GeV (middle) and $300$ GeV (bottom).
 Right: Predicted annihilation cross section for $\chi\chi\to b\overline{b}$ 
 at late times, relevant annihilation in the Milky Way, versus $m_\chi$.
 Purple curves are fixed by the relic density, while 
 the red and green shading denote the $2\,\sigma$ allowed regions \cite{Cholis:2019ejx} for
 gamma-ray and antiproton excesses, respectively. 
 The blue and orange regions are excluded by the Higgs invisible
 width and gamma-ray observations from dSphs~\cite{Fermi-LAT:2016uux, Ando:2016ang}, respectively.}
\label{fig1}
\end{figure*}

We performed a numerical scan of the parameter space, to 
look for regions compatible with the gamma-ray and antiproton
excesses, as well as DM relic abundance and dwarf spheroidal 
constraints on DM annihilation.
The results  are shown in Fig.~\ref{fig1}, where
the extra Higgs boson mass takes the values
 $m_{h_2}=70,\,96,\,300~\mathrm{GeV}$ in successive rows.

In the left panels, the purple lines show the relation between
$\lambda_S$ and $m_\chi$
that gives the correct DM relic abundance, for several values of
the mixing angle, $\st=0.01$ (solid), $0.10$ (dashed), $0.30$ (dot-dashed).
The upper gray and blue regions are excluded by the 
perturbative unitarity bound
$\lambda_S<8\pi/3$~\cite{Chen:2014ask}, and the Higgs invisible decay
constraint $\mathrm{BR}_\mathrm{inv}<0.19$~\cite{Sirunyan:2018owy}, 
respectively.
{We remark that the relic density curves
 may be affected by early
kinetic decoupling of the DM if its mass is just below the Higgs 
resonance
($m_\chi\lesssim m_h/2$)~\cite{Binder:2017rgn, Duch:2017nbe}. As a
result, the Higgs invisible decay constraint may
change by a factor $\lesssim 7$~\cite{Hektor:2019ote}. However this does
not impact the region of parameter space relevant for 
explaining the cosmic ray anomalies, so we do not take account of this
effect.}

In the right panels of Fig.~\ref{fig1}, the relic abundance curves
(purple) are replotted in the plane of the DM mass and the
$s$-wave annihilation
cross section, evaluated at the present time, which is relevant for
indirect detection.  Along each curve, the quartic coupling $\lambda_S$
(or equivalently $v_s$) is fixed to give the correct relic abundance.
To fit the cosmic ray excesses, we adopt the 2$\,\sigma$ allowed 
regions found in Ref.~\cite{Cholis:2019ejx}, plotted as the 
red and green regions for the gamma ray and antiproton signals,
respectively.  For each of the three chosen $m_{h_2}$ values,
there is some overlap between all the allowed regions,
roughly for DM masses in the range 
$64~\mathrm{GeV}\lesssim m_\chi\lesssim67~\mathrm{GeV}$.
The regions excluded by invisible Higgs decays can be inferred from 
the corresponding left-hand graph by noticing where the relic density
curve goes above the Higgs decay curve for a given mixing angle.  It
happens that these excluded regions nearly coincide in mass,
independently of $\theta$, leading to only a single excluded region
in the right-hand plots.

For all but the lowest value of $m_{h_2}$,
the allowed regions are independent of the mixing angle $\theta$,
because only the combination $\st/v_s$ appears in the
$b\overline{b}$ annihilation cross section given by Eq.\ 
(\ref{sigv}).  
For $m_{h_2}=70$\,GeV however, the additional annihilation channel
$\chi\chi\to h_2 h_2$ becomes important, breaking this degeneracy.
We emphasize that in generic models such as singlet scalar DM, 
the allowed region for explaining both excesses is excluded by the strong
direct detection bound, or suppression of the indirect signal.

The allowed regions are also compatible with the most recent 
constraints from DM annihilation in satellite galaxies of the Milky
Way~\cite{Fermi-LAT:2016uux, Ando:2016ang}, shown in orange. This exclusion is based
on gamma-ray observations of 28 confirmed dwarf spheroidal galaxies
(dSph) and $17$ new candidate systems, taking advantage of 
spectroscopically determined $J$-factors for the known dSphs and
and predicted $J$-factors for the candidates.  It gives a limit that
is 1.5 times weaker than in previous determinations.

In our numerical analysis, only the annihilation channel $\chi\chi\to
b\overline{b}$ is taken into account for the gamma-ray and antiproton
excesses.  However for heavy DM, the annihilation into $WW^*$ may
dominate. We have checked that for $m_\chi$  in the range
$55~\mathrm{GeV}\lesssim m_\chi\lesssim65~\mathrm{GeV}$, the branching
fraction into $WW^*$ can be $25\%$ at most, and the ensuing
shapes of the energy spectra of gamma rays and antiprotons are only
slightly modified as a result.

\section{Potential stability and perturbativity}
\label{sec:stability}

Another interesting feature of our model is that the allowed values of the
Higgs portal coupling $\lambda_{HS}$ can be compatible with extending the
stability of the Higgs potential up to the Planck
scale~\cite{Lebedev:2012zw, EliasMiro:2012ay,Khoze:2014xha,Falkowski:2015iwa,Athron:2018ipf}. It is
well-established that the Higgs quartic coupling in the SM runs to negative
values at high energy scales $\Lambda_I\sim 10^{11}$\,GeV under
renormalization~\cite{Degrassi:2012ry}.  This could lead to a
catastrophic destabilization 
of the electroweak vacuum in the early universe, if inflation occurred at a
sufficiently high scale~\cite{East:2016anr}.

There are two ways in which the Higgs portal coupling $\lambda_{HS}$ can 
improve stability. First, it provides an additional positive contribution to
the $\beta$ function of $\lambda_H$. Second, $\lambda_H$ gets a positive
threshold correction at the scale $\mu = m_{h_2}$ when $h_2$ is integrated
out. The running $\lambda_H$ is increased by $\delta\lambda_H \equiv
\lambda_{HS}^2/\lambda_S$ as the scale $\mu$ crosses the threshold from below.   For
$\lambda_{HS}>0$, the stability condition also becomes more stringent  at
this scale, $\lambda_H > \delta\lambda$ instead of $\lambda_H>0$.  But as
the scale $\mu$ becomes sufficiently large compared to $m_{h_2}$, the condition
relaxes to the naive requirement $\lambda_H(\mu) > 0$, so that the threshold
correction can overcome the tendency for $\lambda_H(\Lambda_I)$ to become
negative.
The $\beta$ functions of the couplings at two-loop level are obtained
using \texttt{SARAH}~\cite{Staub:2010jh, Staub:2013tta}. 
Here we concentrate on the $\lambda_{HS}>0$ case since this has been
shown to be more effective for Higgs stability than 
$\lambda_{HS}<0$ \cite{EliasMiro:2012ay}.

\begin{figure}[t]
\begin{center}
 \includegraphics[scale=0.7]{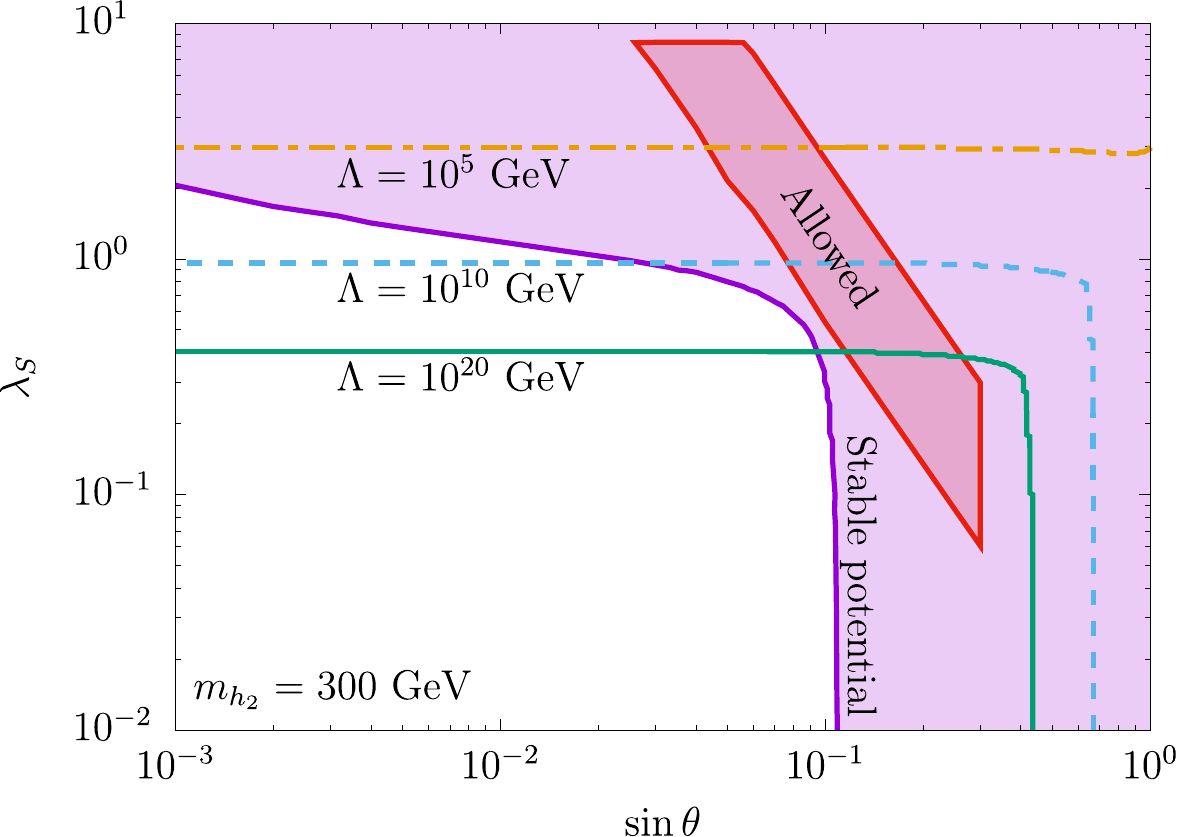}
  \includegraphics[scale=0.7]{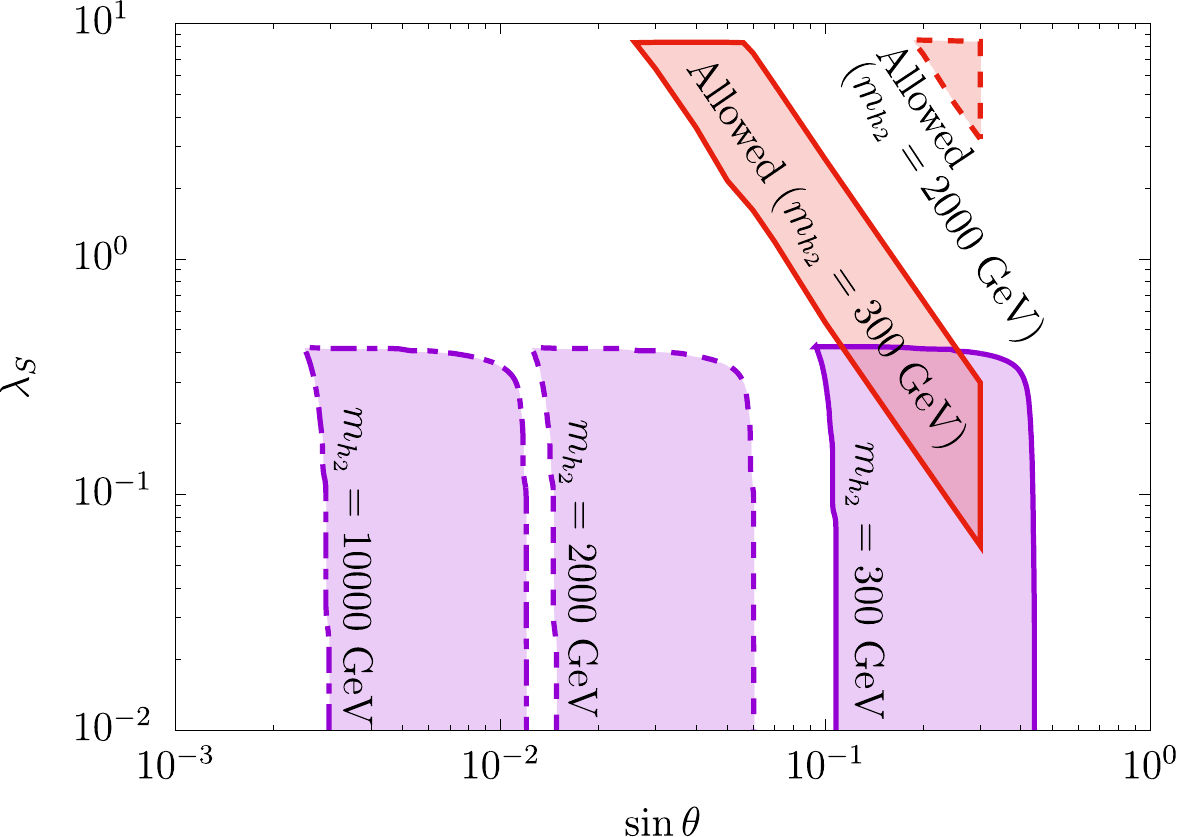}
 \caption{Top: regions $\lambda_S$ versus $\sin\theta$ where the 
 scalar potential is
 stable and perturbative, taking
 $m_{h_2}=300~\mathrm{GeV}$. In the purple region, the 
Higgs quartic
coupling $\lambda_H$ is stabilized by the contributions of $\lambda_{HS}$
 to its running under the renormalization.
 The red region can explain the cosmic ray anomalies consistently the
 constraints considered in the previous section. 
 The other curves
(solid green, dashed light blue, dot-dashed orange) represent the upper limit on $\lambda_S$
or $\sin\theta$ to avoid a Landau pole near the renormalization scale
 indicated.
 Bottom: region that the scalar potential is stable and all the quartic
 couplings are perturbative up to the Planck scale (purple region) for
 $m_{h_2}=300$, $2000$, $10000$ GeV. The red region is the same with the
 upper plot, but for $m_{h_2}=300$ GeV (solid) and $2000$ GeV (dashed).}
 \label{fig:rge}
\end{center} 
\end{figure}

In Fig.~\ref{fig:rge} we display the regions  $\sin\theta$-$\lambda_S$
plane that correspond to models with Higgs stability up to the Planck
scale (purple region) and that can explain the cosmic ray excesses
(red region).   The upper plot illustrates the situation for
$m_{h_2}=300$ GeV. The regions above and to the right of the curves
labeled by $\Lambda = 10^5$, $10^{10}$, $10^{20}$ GeV have a Landau pole in one of the
quartic couplings $\lambda_H$, $\lambda_S$ and $\lambda_{HS}$ at the
scale indicated. Thus a fully UV-complete theory can be achieved for
parameters in the range $0.1\lesssim\sin\theta\lesssim 0.3$ and
$0.06\lesssim\lambda_S\lesssim 0.4$. The threshold correction, rather
than the extra contribution of  $\lambda_{HS}$ to the $\beta$
function, is the most important effect for achieving  stability in the
overlap region.  For $\sin\theta\gtrsim0.4$, it gives too large an
enhancement in $\lambda_H$, causing it to blow up at scales $\Lambda$
below the Planck scale; this is the reason the Landau pole
curves are vertical at large $\sin\theta$.

In the lower plot of Fig.~\ref{fig:rge}, we show the allowed
parameters corresponding to higher values of $m_{h_2}$.  Here the purple shaded
regions combine the requirements of stability of the Higgs potential
and perturbativity of the quartic couplings, considering 
$m_{h_2}=300$, $2000$ and $10^4$ GeV, requiring smaller
$\sin\theta$ with increasing $m_{h_2}$.  The allowed region for
the cosmic ray anomalies shrinks rapidly with increasing $m_{h_2}$.
For $m_{h_2}\gtrsim600$ GeV, the overlap of the stability and
anomaly regions disappears.  Moreover for 
$m_{h_2}\lesssim140~\mathrm{GeV}$, there is no longer any 
region of stability/perturbativity~\cite{Falkowski:2015iwa}.

\section{LEP and CMS anomalies}
\label{sec:lep-cms}

There are experimental hints for an additional Higgs-like state
with mass near 96\,GeV.  In their search for the SM Higgs,
the combined LEP collaborations observed 
a 2.3\,$\sigma$ excess of $b\bar b$ final states at $m\sim 98$\,GeV,
corresponding to a signal strength approximately 10 times lower than
would be produced by a SM Higgs of this mass~\cite{Barate:2003sz}.  

More recently, in a search for the diphoton final state, CMS reported a
2.9\,$\sigma$ (local significance) excess in the 13\,TeV data at $m =
95.3$\,GeV, with a production cross section times branching ratio of
$0.13$\,pb, close to that which a SM Higgs of the same mass would
produce~\cite{CMS:2015ocq,CMS:2017yta}. 
To simultaneously accommodate both anomalies, we adopt the values
taken in ref.~\cite{Heinemeyer:2018wzl}
\bea
	{\sigma(e^+e^-\to h_2\to Z b\bar b)\over
	\sigma(e^+e^-\to h\to Z b\bar b)} &=& \mu_{\rm\sss LEP}=0.117\pm0.057\nn\\
	{\sigma(gg\to h_2\to\gamma\gamma) \over
	\sigma(gg\to h\to\gamma\gamma)} &=& \mu_{\rm\sss CMS}=0.6\pm 0.2 
\label{lep-cms}
\eea
where $h$ denotes a SM-like Higgs with mass 96\,GeV.  

Although the LEP anomaly could be explained by Higgs mixing with angle
$\sin\theta \sim \pm\sqrt{0.117} = \pm 0.34$, this would not be sufficient to
produce the CMS excess.  As pointed out in ref.~\cite{Fox:2017uwr},
one needs to couple $h_2$ to new charged states to enhance the
branching ratio (by a factor of $\sim 7$) for $h_2\to\gamma\gamma$ via the
one-loop diagram containing the exotic states.

\subsection{Adding charged scalars}

We investigated models with charged colored scalars $\Phi$ that decay
into two or four quarks.   One might think that the latter case is
 more difficult to probe at the LHC, since the signal would be
pair production of $\Phi$ followed by decays into four jets each.
However this mode has recently been tightly constrained by
CMS~\cite{Sirunyan:2018zyx}, with $m_{\Phi} > 710\,$GeV (1.4\,TeV) for
colored triplets (octets).
It turns out that models where $\Phi$ can decay into two jets are less
stringently constrained at present~\cite{Aaboud:2017nmi}.  One could also
consider models where $\Phi$ decays to six quarks through a dimension-10 operator.
However we estimate that the scale $\Lambda$ suppressing such an operator must
be $\lesssim 3\,$TeV to avoid $\Phi$ exiting the detector before decaying, and being
identified as a long-lived charged particle, which is also excluded for $m_\Phi \lesssim
1\,$TeV for color triplets~\cite{Khachatryan:2016sfv}.  We do not consider 
this case in the following.

\begin{table}[]
\centering
\tabcolsep 2.5pt
\begin{tabular}{|l|c|c|c|c|c|c|c|c|}
\hline
model& $q_\Phi$ & $N_c$ & 
${m_{\Phi}\over|\lambda_{S\Phi}|^{1/2}}$ &
${\bar \mu_{\Phi}\over|\lambda_{S\Phi}|^{1/2}}$
& $\st$ & ${\lambda_{S\Phi}}$ & ${\lambda_{H\Phi}}$ & 
 $\nicefrac{\chi^2}{{\rm d.o.f.}}$\\
\hline
1 & $\nicefrac83$ & 6 &  943 & 836 & 0.39 & 1.9 & 3.3 & $3.6$ \\
2 & $\nicefrac83$ & 3 &  601 & 778 & 0.36 & 1.4 & 1.6 & $2.2$\\
3 & $\nicefrac53$ & 6 &  700 & 741 & 0.34 & 3.4 & 3.5 & $2.1$ \\
4 & $\nicefrac53$ & 3 &  417 &  838 & 0.39 & 3.0 & 5.2 & $3.7$\\
\hline
5 & $\nicefrac23$ & 6 &  588 &  795 & 0.37 & 4.8 & 5.9 & $1.4$\\
6(*) & $\nicefrac23$ & 3 &  284 &  765 & 0.35 & 3.4 & 3.6 & $1.5$\\
7 & $\nicefrac{-1}3$ & 6 &  554  & 830 & 0.39 & 5.4 & 8.0 & $1.5$\\
8(*) & $\nicefrac{-1}3$ & 3 &  256  & 810 & 0.38 & 4.1 & 5.6 & $1.4$\\
\hline
9 & $\nicefrac{-4}3$ & 6 &  666 & 752 & 0.35 & 3.8 & 3.9 & $1.8$\\
10(*) & $\nicefrac{-4}3$ & 3 &  333  & 737 & 0.34 & 2.4 & 3.0 & $2.5$\\
\hline
\end{tabular}
\caption{Models of SU(3)$_c$ sextet or triplet scalars coupling to $S$,
considered for explaining the CMS diphoton excess.  
The mass $m_\Phi$ (in GeV) and mixing
angle are chosen to fit the combined LEP and CMS anomalies.
$\bar \mu_\Phi$ represents the contribution of the $S$ and $H$ 
VEVs to the mass; see Eq.~(\ref{phimass}).
The ratio of couplings $\lambda_{H\Phi}/\lambda_{S\Phi}$ 
is chosen to mitigate Higgs coupling strength deviations,
leading to the reduced $\chi^2$ values in the last column.
Models marked with $(*)$ can optionally refer to the
two-quark coupling version,
while those without asterisk couple $\Phi$ to four quarks.
The magnitude of $\lambda_{S\Phi}$ is such that $m_\Phi$
saturates the experimental constraints on $\Phi$ decaying to
$qqqq$~\cite{Sirunyan:2018zyx} (or $qq$~\cite{Sirunyan:2018rlj} for the models
marked with ``*'').
\label{tab0}}
\end{table}

\begin{table*}[]
\centering
\begin{tabular}{|c|c|c|c|c|c||c|}
\hline
observable &${gg\to h_1\atop \to\gamma\gamma}$ & 
${gg\to h_1\to\atop ZZ \to 4\ell}$ &
$\kappa_g$ & $\kappa_\gamma$ & 
$\kappa_{Z,W}$ &$\langle\sigma_{\gamma\gamma}v_\mathrm{rel}\rangle$
($10^{-28}$\,cm$^3$/s)\\
\hline
predicted & $\kappa_g^2\,c_{\gamma\gamma}$ & $\kappa_g^2\,c_{ff}$ &
$|\sfrac32 b_1^g|$ & $|b_1^\gamma/b_{\rm SM}^\gamma|$ & $c_\theta $ 
&  Eq.\ (\ref{sigvgg})\\
\hline
model 1 & 0.87 & 0.82 & 0.91 & 0.95 & 0.92 & $0.9$\\
model 2 & 1.02 & 0.83 & 0.91 & 1.03 & 0.93 & $1.1$ \\
model 3 & 0.91 & 1.00 & 1.01 & 0.89 & 0.94 & $0.5$ \\
model 4 & 0.87 & 0.82 & 0.90 & 0.95 & 0.92 & $0.9$ \\
\hline
model 5 & 1.03 & 1.06 & 1.04 & 0.92 & 0.93 & $0.6$\\
model 6 & 0.99 & 1.04 & 1.03 & 0.91 & 0.94 & $0.6$\\
model 7 & 1.03 & 1.03 & 1.03 & 0.92 & 0.92 & $0.6$\\
model 8 & 1.02 & 1.04 & 1.03 & 0.92 & 0.93 & $0.6$\\
\hline
model 9 & 0.95 & 1.02 & 1.02 & 0.90 & 0.94 & $0.5$\\
model 10 & 0.88 & 0.94 & 0.97 & 0.91 & 0.94 & $0.5$\\
\hline
CMS & ${1.15\pm 0.15}$~\cite{CMS:1900lgv}& 
${0.94 \pm 0.10}$~\cite{CMS:2019chr} &
$1.18{+0.16\atop-0.14}$~\cite{Sirunyan:2018koj}& 
${1.07\pm 0.15}$~\cite{Sirunyan:2018koj}& 
$\kappa_Z={1.00\pm 0.11}$~\cite{Sirunyan:2018koj} &  Fermi/LAT:\\
ATLAS & $0.96\pm 0.14$~\cite{ATLAS:2019slw} & $1.04{+0.16\atop-0.15}$~\cite{ATLAS:2019slw}
&$0.99{+0.11\atop-0.10}$~\cite{ATLAS:2019slw}  &
$1.05\pm0.09$~\cite{ATLAS:2019slw}& 
${\kappa_W=1.05\pm 0.09\atop
\kappa_Z=1.11\pm0.08}$~\cite{ATLAS:2019slw} & 
$\lesssim (0.5-4)$~\cite{Ackermann:2015lka}\\
\hline
combined & $1.06\pm 0.10$ & $0.99\pm 0.10$ & $1.09{+0.10 \atop -0.09}$&
$1.06 \pm 0.09$ & $1.05\pm  0.08$  & $< 1.4$\\
\hline
\end{tabular}
\caption{Predicted SM Higgs signal strengths, and observed values,
for the 10 models considered. Last column shows the cross section for
$\chi\chi\to\gamma\gamma$ annihilation in the galaxy, and at the
bottom the limit
\cite{Ackermann:2015lka} assuming Einasto profile for the Milky Way DM halo.
\label{tab1}}
\end{table*}

The models, which are summarized in Table~\ref{tab0},
are represented by the interaction Lagrangian
\bea
	{\cal L} &\ni& \lambda_{S\Phi}  |S|^2|\Phi|^2 + 
	\lambda_{H\Phi}  |H|^2|\Phi|^2 \nn\\
	&+& \left(y_\Phi\, \Phi^* (\bar q_\R q_\R^c) \hbox{\ \  or \ }
	 {1\over\Lambda^3}\Phi\,(\bar q_\R q_\R^c)^2 \right) +{\rm H.c.}
\label{Lint}
\eea
In the last term, a dimension-7 effective operator, the right-handed
quarks are assumed to form a sextet or triplet of color, the smallest
possible irreducible representations, 
which can combine
with $\Phi$ to form a singlet.  The possible flavor combinations
are $qqqq = uuuu,$ $uuud$, $uudd$, $uddd$ and $dddd$ corresponding
to charges $q_\Phi = \nicefrac83$, $\nicefrac53$, $\nicefrac23$, $-\nicefrac13$ and $-\nicefrac43$,
respectively, where $u$ and $d$ refer to generic up- and down-type
quark flavors.   For the two-quark models, $\Phi$ could transform as
a $\bar 6$ or a 3 of SU(3)$_c$, but we consider only the color triplet case, which is less
strongly constrained by collider limits,
with $qq= uu$, $ud$, $dd$ and corresponding charges $q_\Phi = \nicefrac43,\,\nicefrac13,\,
-\nicefrac23$.

\begin{figure}[b]
\begin{center}
 \includegraphics[scale=0.6]{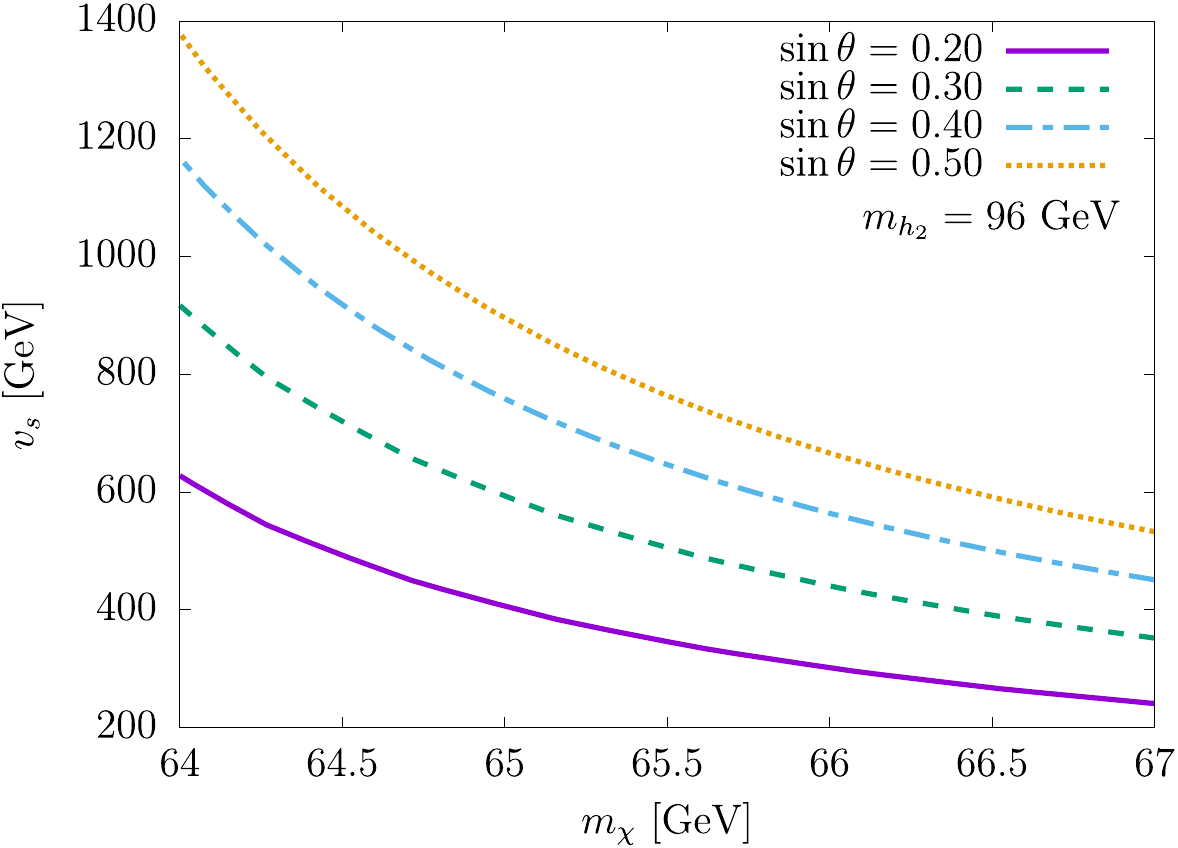}
 \caption{The singlet VEV $v_s$ versus DM mass $m_\chi$ required by
the relic abundance, for several choices of mixing angle, assuming
$m_{h_2} = 96\,$GeV.}
 \label{fig:vs}
\end{center} 
\end{figure}

The 
first two interactions in (\ref{Lint}) produce a shift in the mass of $\Phi$ when
$S$ and $H$ get their VEVs.  If $\mu^2_\phi$ denotes the original Lagrangian
parameter, then the physical mass is given by
\bea
	m_\Phi^2 &=& \mu_\Phi^2 + \sfrac12\lambda_{S\Phi} v_s^2 +
	\sfrac12\lambda_{H\Phi} v^2 \nn\\
	&\equiv& \mu_\Phi^2 + \bar \mu_\Phi^2
\label{phimass}
\eea
One would like to avoid having $m_\phi^2 \ll |\mu^2_\phi|$ or $|\bar
\mu_\phi^2|$ since this would require fine-tuning; Table~\ref{tab0}
shows that some mild tuning ($\sim 0.1$) is needed for models 6 and 8.

\subsection{Couplings to $g,\,\gamma$, $h_1$}

The effective couplings of $h_{1,2}$ to gluons and photons have been computed
for a similar class of models~\cite{Nakamura:2017irk},
\bea
	{\cal L_{\rm eff}} =\sum_{i=1}^2\left( {\alpha_\mathrm{s} b_i^g\over 8\pi v}
	 \,h_i\, G_{\mu\nu}^a G^{a\mu\nu} + 
	{\alpha_\mathrm{em}  b_i^\gamma\over 8\pi v} \,h_i \,
	F_{\mu\nu} F^{\mu\nu}\right),\qquad
\eea
with
\bea
	b_1^\gamma &=& {N_c\over 3} q_\Phi^2(-\st\eta_S +\ct\eta_H) + 
	b_{\rm SM}^\gamma\ct\nn\\
	b_1^g &=& R_\Phi(-\st\eta_S +\ct\eta_H) + \sfrac23\ct\nn\\
	b_2^\gamma &=& {N_c\over 3} q_\Phi^2(\ct\eta_S +\st\eta_H) + 
	b_{\rm SM}^{\gamma '}\st\nn\\
	b_2^g &=& R_\Phi(\ct\eta_S +\st\eta_H) + \sfrac23\st
\eea
Here $N_c=6\,(3)$ is the number of colors of $\Phi$,  
$R_\Phi = \nicefrac56\,(\nicefrac16)$
for sextet (triplet) scalars, $b^\gamma_{\rm
SM}\cong -6.5$, $b^{\gamma '}_{\rm
SM}\cong-5.9$,\footnote{This is somewhat smaller than the value for
the SM Higgs, due to $m_{h_2}$ being less than $m_{h_1}$.
$R_\Phi$ is equal to the index of the representation, divided by 6 for
a complex scalar field.}\  
$\eta_S = \lambda_{S\Phi} v v_s/ m_\Phi^2$ and
$\eta_H = \lambda_{H\Phi} v^2/ m_\Phi^2$.  The value of
$v_s$ giving the correct relic density for
$m_{h_2}=96\,$GeV is  plotted in Fig.~\ref{fig:vs}, and can be fit to
$v_s \cong (-35 + 3641\,\st - 1603\,\st^2)\,$GeV
at $m_\chi = 64\,$GeV, while reaching smaller values 
$v_s\cong 250\sim550$\,GeV at
$m_\chi = 67$\,GeV.  Since the solution for $m_\Phi$ derived below is
proportional to $v_s^{1/2}$, in order to maximize the scale of new physics
we adopt the lower value of $m_\chi$, which is consistent with 
the cosmic ray anomalies as described previously.

\begin{figure}[t]
\begin{center}
 \includegraphics[scale=0.7]{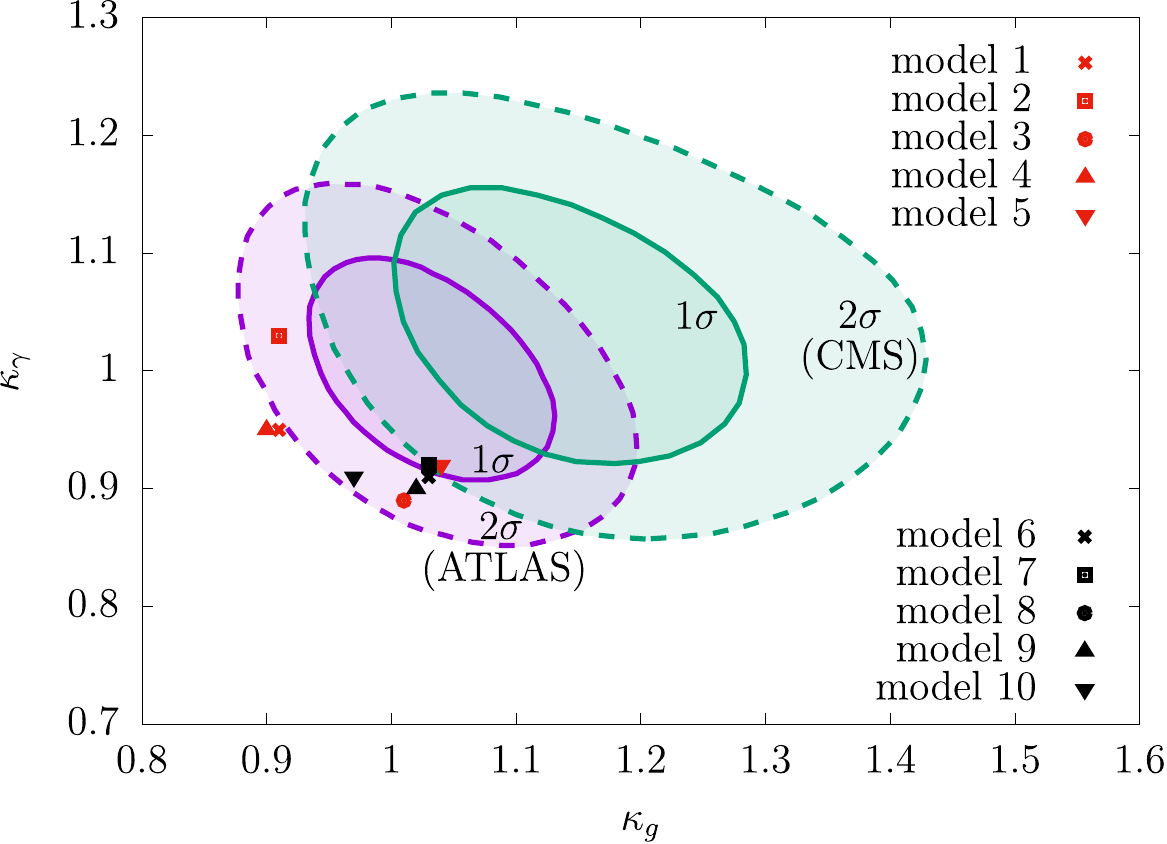}
 \caption{Effective Higgs coupling strengths to photons versus
gluons predicted by the 10 models, and the allowed regions from 
ATLAS~\cite{ATLAS:2019slw} and CMS~\cite{Sirunyan:2018koj}.}
 \label{fig:kk}
\end{center} 
\end{figure}

The couplings of the SM Higgs to gluons and photons get modified by
mixing and the $\Phi$ loop.  Couplings of $h_1$ to fermions and
$W/Z$ bosons are reduced by the mixing effect alone.   This can be
quantified by 
$\kappa_i$ parameters, defined as the 
coupling to particle $i$ divided by the corresponding
value in the SM:
\bea
   \kappa_{g} = {b_1^g\over \nicefrac23},\quad
   \kappa_{\gamma} = {b_1^\gamma \over b^\gamma_{\rm
SM}},\quad
	\kappa_{\sss W,Z} = \kappa_f = \ct\ .
\label{kappas}
\eea
These factors appear in the signal strengths for various Higgs
production and decay processes.  In addition we make use of the 
ratio of branching ratios of the Higgs to different final states,
relative to the corresponding SM values.  If there was only mixing,
these would be trivial, but because of the $\Phi$ loop, the $gg$
and $\gamma\gamma$ channels can change relative to the others.
These factors are
\bea
	c_{ff} &=& {1\over \BR(f)+ \BR(g)\, \mu_g/\ct}\nn\\
	c_{\gamma\gamma} &=& {\mu_\gamma\over \BR(f)\,
	\ct^2 +\BR(g)\,\mu_g}
\eea
where BR($g) = 0.086$ is the SM branching ratio into gluons
at $m_h=96\,$GeV,
and BR($f) = 0.914$ is that into other states.
 Then for example the signal strength for $gg\to
h\to\gamma\gamma$ is $\kappa_g^2 c_{\gamma\gamma}$, while that for
$gg\to h\to ZZ\to 4\ell$ is $\kappa_g^2 c_{ff}$.

\subsection{Fitting the anomalies}

By taking the square root, Eqs.~(\ref{lep-cms}) can be recast in the form
\bea
\label{lep-cms2a}
	\st^2 &=& \sqrt{\mu_{\rm\sss LEP}\,\hat\Gamma_{h_2}}\\
	\left({b_2^g\over \nicefrac23}\right)
	\left({b_2^\gamma \over b^{\gamma '}_{\rm SM}}\right)
	&=& \pm \sqrt{\mu_{\rm\sss CMS}\,\hat\Gamma_{h_2}}
\label{lep-cms2b}
\eea
where $\hat\Gamma_{h_2} = \BR(f)\st^2 + \BR(g)(\sfrac32 b_2^g)^2$
is the ratio of the $h_2$ width to that of a SM-like Higgs of the
same mass.  For fixed values of $\lambda_{H\Phi}$ and  
$\lambda_{S\Phi}$, these can be solved numerically for
$\theta$ and $m_\Phi$.  The equations are only weakly coupled,
through the small gluon contribution to $\hat\Gamma_{h_2}$; therefore
$\st$ is practically fixed by~(\ref{lep-cms2a}), while $m_\Phi$
is mostly determined by~(\ref{lep-cms2b}).

For solutions of (\ref{lep-cms2a}) and (\ref{lep-cms2b}), 
the Higgs couplings~(\ref{kappas}) depend
upon the ratio $\eta_{H}/\eta_{S} =
v\lambda_{H\Phi}/(v_s\lambda_{S\Phi})$, which we determine by
fitting to the set of signal strengths and effective couplings
shown in Fig.~\ref{fig:kk}.  The $\chi^2$ per degree of freedom
(five observations from combining ATLAS and CMS results,
 minus one parameter) is listed in Table~\ref{tab0}, and the
results for $\kappa_\gamma$ versus $\kappa_g$ are illustrated in Fig.~\ref{fig:kk}.
There is a preference for models 5-8, in which the scalar
has charge $q_\Phi=\nicefrac23$ or $-\nicefrac13$, and can be either a triplet or
a sextet.  Models 2,\,3 with $q_\Phi = \nicefrac83,\,\nicefrac53$ and $N_c = 3,6$
respectively also give an acceptable fit, as well as model 9 with $q_\Phi=-\nicefrac43$,
$N_c=6$.

The models face daunting constraints from LHC searches
for pair production of colored scalars that decay into jets.
Gluinos must be heavier than $1.5\,(1.4)\,$
TeV~\cite{Aaboud:2017nmi,Sirunyan:2018zyx} for decays into 2\,(4) jets.\footnote{We have compared
the leading-order production cross section for gluinos \cite{Beenakker:1995fp} to that of color sextet scalars 
\cite{Chen:2008hh}
at the same mass and
$\sqrt{s}=13\,$TeV, averaged over parton distribution functions \cite{Martin:2009iq},
and found that they are quite similar, suggesting a bound of 1.3\,TeV for 
decays into four jets.}
Searches for $R$-parity violating top squark 
decays by CMS~\cite{Sirunyan:2018rlj} and ATLAS~\cite{Khachatryan:2014lpa,Veeraraghavan:2016nqk,Aaboud:2017nmi,ATLAS:2017qih}
put complementary
constraints on triplets decaying to two jets, that depend upon 
whether one of the jets is tagged for $b$ quarks: $m_{\tilde t} > 520\,$GeV
for decays into first or second generation quarks $qq$, versus $m_{\tilde t} > 610$\,GeV
for decays into $bq$.

Table~\ref{tab1} shows that one must take rather large values of the couplings
$\lambda_{S\Phi},\,\lambda_{H\Phi}$ to satisfy the collider limits on $m_\Phi$.
Although these values are consistent with perturbative unitarity, they typically
require some fine tuning since the one-loop contributions to $\lambda_H$, $\lambda_S$
and $\lambda_{HS}$ are of order
\bea
	\delta\lambda_H &\sim& {3\lambda_{H\Phi}^2\over 16\pi^2},\quad
	\delta\lambda_{HS} \sim {3\lambda_{H\Phi}\lambda_{S\Phi}\over 16\pi^2},\quad
	\delta\lambda_S \sim {3\lambda_{S\Phi}^2\over 16\pi^2}\nn\\
\eea
which tend to be much larger than 
 the tree-level values $\lambda_H\sim 0.24$, $\lambda_S\sim 0.009$
and $\lambda_{HS}\sim 0.008$.  Model 2 however is relatively good in this
respect, with $\delta\lambda_H\sim \delta\lambda_{HS}\sim \delta\lambda_S \sim 0.04$,
requiring only a mild 1 part in 5 tuning.

The $\Phi$-extended model needed for the collider anomalies, although not ruled out, is tightly
constrained, and would likely require other new physics to be part of a UV-complete picture. 
For example, stability of the Higgs potential is not helped by the small $\lambda_{HS}$
coupling in these examples, although the much larger $\lambda_{H\Phi}$ coupling may do so. 
Pending experimental verification, we content ourselves with the foregoing low-energy description as
concerns the tentative LEP/CMS excesses.

\subsection{Gamma-ray line searches}

The Fermi-LAT has obtained stringent constraints on DM
annihilating to two monochromatic photons~\cite{Ackermann:2015lka}, which occurs through
the Higgs portal in our model, and is potentially enhanced by the
$\Phi$ induced coupling of photons to the singlet.  The cross
section for $\chi\chi\to\gamma\gamma$ by $h_{1,2}$ exchange in the
$s$-channel is
\begin{equation}
 \langle\sigma_{\gamma\gamma}v_\mathrm{rel}\rangle \approx
  \frac{\alpha_\mathrm{em}^2s}{128\pi^3 v^2v_s^2}
  \left|
   \frac{s_\theta m_{h_1}^2b_1^{\gamma}}{s-m_{h_1}^2}
   -\frac{c_\theta m_{h_2}^2b_2^{\gamma}}{s-m_{h_2}^2}
  \right|^2
  \label{sigvgg}
\end{equation}
where $s = 4m_\chi^2$ and the decay width in the $s$-channel propagators
were neglected.\footnote{There is also a contribution from direct
coupling of $\chi\chi$ to $\gamma\gamma$ through the $\Phi$ loop,
whose matrix element is $|{\cal M}| = \alpha_\mathrm{em} q_\Phi
N_c\lambda_{S\Phi} q_\Phi^2 m_\chi^2/(3\sqrt{2}\pi\, m_\Phi^2)$ 
\cite{Cline:2012nw}, which is subdominant to the $h_{1,2}$-mediated
contributions.}\ \   (Here we modify the definition of $b_2^\gamma$
to be in terms of $b_\mathrm{SM}^\gamma$ rather than $b_\mathrm{SM}^{\gamma'}$ 
since the same energy $\sqrt{s}$ is flowing through the loops coupling to $h_1$
and $h_2$.)  In the last column of Table~\ref{tab1} we show the
predictions for the $\Phi$-extended models. 
The corresponding prediction for the minimal model with
no scalar $\Phi$ can be obtained from this by setting $N_c=0$ in
$b_{1,2}^\gamma$, and can always be made parametrically small by
taking $\st \ll 1$, unlike in the extended models.  We find that the
allowed regions shown in previous sections are not reduced by
this constraint.

The gamma-ray line constraints on $\langle\sigma v\rangle_{\gamma\gamma}$
become stronger or weaker depending on how cuspy one assumes the
density profile of DM is in the Milky Way.  Therefore
ref.~\cite{Ackermann:2015lka} considers four possibilities of varying 
cuspiness, resulting in a range of upper bounds shown in
Table~\ref{tab1}.
For reference we adopt an intermediate upper limit
$\langle\sigma_{\gamma\gamma}v_\mathrm{rel}\rangle<1.4\times10^{-28}~\mathrm{cm^3/s}$
at $m_\chi=64$ GeV, based on the assumption of an Einasto DM profile.

Interestingly the predicted values for $\langle\sigma
v\rangle_{\gamma\gamma}$ are not far below the averaged upper limit, leading
to the expectation that an observable gamma-ray line from the
galactic center, of 
energy $\sim 64\,$GeV, should be correlated with the LEP and CMS
anomalies.
The predicted line will be tested by the next generation
gamma-ray telescope GAMMA-400~\cite{Galper:2018gku}.

\section{Conclusion}
\label{conclusion}

Annihilation of $\mathcal{O}(60)~\mathrm{GeV}$ mass dark matter
into $b$ quarks remains as a plausible explanation for excess cosmic
gamma-rays and antiprotons that have been observed by Fermi/LAT and
AMS respectively.  We have shown that an economical model of scalar
DM coupling to the Higgs can naturally explain these
observations, while respecting other constraints.  In particular, the
pseudo-Goldstone nature of pNGB DM makes it immune to direct
detection, because of its highly velocity-suppressed couplings to
nucleons.

We showed that, while firmly within the preferred parameter space
for  antiprotons, our prediction for the DM mass $m_\chi \cong
(64-67)\,$GeV is at the high end of the $2\,\sigma$-allowed region
for the galactic center gamma-ray excess, and the annihilation cross
section is not far below the Fermi indirect detection limit from
dwarf spheroidal galaxies.  Further, if the extra
$CP$-even Higgs boson is in the mass range $m_{h_2}\sim (200-600)$\,GeV,
this allowed region can overlap with parameters for which the
Higgs potential remains stable up to the Planck scale, and the new
quartic couplings are free from Landau poles.

By extending the model with an additional scalar $\Phi$ that 
carries charge and color, and assuming that $m_{h_2}\sim 96\,$GeV,
we can also account for tentative excess $b$ quarks 
observed at LEP and diphotons at CMS.  The most promising such model
has $\Phi$ in the $3$ representation,
$m_\Phi\gtrsim 720\,$GeV, charge $q_\Phi = \nicefrac83$, and decaying
into four up-type quarks in the first or second generations.
It can eventually be ruled out by strengthened limits on pair 
production of such scalars, or by more precise observations of the
production and decay modes of the standard model Higgs boson,
with which it has some tensions at the $2\,\sigma$ level.
A further test is through the search for a monochromatic gamma-ray
line at $\sim 64\,$GeV from the galactic center, which could rule
out the extended model with a modest increase in sensitivity over
current limits.  This extension would require some other new physics
to stabilize the Higgs self-coupling and the new scalar cross-couplings
at high scales.

Higgs sectors extended by couplings to singlet fields have been
invoked to strengthen the electroweak phase transition, possibly
enabling electroweak baryogenesis or producing observable gravity
waves.  This is one feature that the pNGB DM does not have, however,
as was shown in ref.~\cite{Kannike:2019wsn}.

\section*{Acknowledgments}
This work was supported by the Natural Sciences and
Engineering Research Council of Canada (NSERC), 
Compute Ontario, WestGrid, Compute Canada, and the Yukawa Institute
Computer Facility.

\end{document}